\documentstyle[amstex,amssymb,12pt,epsfig]{article}
\newlength{\dinwidth}
\newlength{\dinmargin}
\begin{document}

%--------------------------------------------------------------------------

%--------------------------------------------------------------------------
%--------------------------------------------------------------------------

\thispagestyle{empty} \vspace*{1cm} \rightline{Napoli DSF-T-08/2006} %
\rightline{INFN-NA-08/2006} \vspace*{2cm}

\begin{center}
{\LARGE CFT description of the Fully Frustrated XY model and phase diagram
analysis}

{\LARGE \ }

{\large Gerardo Cristofano\footnote{{\large {\footnotesize Dipartimento di
Scienze Fisiche,}{\it \ {\footnotesize Universit\'{a} di Napoli ``Federico
II''\ \newline
and INFN, Sezione di Napoli},}{\small Via Cintia, Compl.\ universitario M.
Sant'Angelo, 80126 Napoli, Italy}}}, Vincenzo Marotta\footnotemark[1] ,
Petter Minnhagen\footnote{{\large {\footnotesize Department of Theoretical
Physics, Umea University}{\footnotesize , 901 87 Umea, Sweden}}}}, {\large %
Adele Naddeo\footnote{{\large {\footnotesize Dipartimento di Fisica {\it ''}%
E. R. Caianiello'',}{\it \ {\footnotesize Universit\'{a} degli Studi di
Salerno \ \newline
and CNISM, Unit\`{a} di Ricerca di Salerno, }}{\small Via Salvador Allende,
84081 Baronissi (SA), Italy}}}, Giuliano Niccoli\footnote{{\large
{\footnotesize Laboratoire de physique,} {\footnotesize Ecole\ Normale\
Sup\'{e}rieure\ de\ Lyon,\ 46\ all\'{e}e\ d'Italie,\ 69364\ Lyon\ cedex\
07,\ France.}}}}.

{\small \ }

{\bf Abstract\\[0pt]
}
\end{center}

\begin{quotation}
Following a suggestion given in \cite{foda}, we show how the $U(1)\otimes
Z_{2}$ symmetry of the fully frustrated XY (FFXY) model on a square lattice
can be accounted for in the framework of the $m$-reduction procedure
developed for a Quantum Hall system at ``paired states''\ fillings $\nu =1$
\cite{cgm2,cgm4}. The resulting twisted conformal field theory (CFT) with
central charge $c=2$ is shown to well describe the physical properties of
the FFXY model. In particular the whole phase diagram is recovered by
analyzing the flow from the $Z_{2}$ degenerate vacuum of the $c=2$ CFT to
the infrared fixed point unique vacuum of the $c=\frac{3}{2}$ CFT. The last
theory is known to successfully describe the critical behavior of the system
at the overlap temperature for the Ising and vortex-unbinding transitions.

\vspace*{0.5cm}


{\footnotesize PACS: 11.25.Hf, 02.20.Sv, 03.65.Fd\newpage }\baselineskip%
=18pt \setcounter{page}{2}
\end{quotation}

\section{Introduction}

A few years ago Foda \cite{foda} proposed that the continuum limit of a FFXY
model is equivalent, at the overlap temperature, to a free massless field
theory of one boson and one Majorana fermion, that is a superconformal field
theory with central charge $c=3/2$. Such a model shares both discrete and
continuous symmetries, which describe well a Josephson junction array in a
transverse magnetic field, provided that the external magnetic flux
threading each cell of the array is $\frac{1}{2}\Phi _{0}$, where $\Phi _{0}=%
\frac{hc}{2e}$ is the superconducting flux quantum \cite{teitel1,halsey1}.

The XY model on a square lattice in the presence of an external magnetic
field transversal to the lattice plane is described by the action:
\begin{equation}
H=-\frac{J}{kT}\sum_{\left\langle ij\right\rangle }\cos \left( \varphi
_{i}-\varphi _{j}-A_{ij}\right) ,  \label{act1}
\end{equation}
where $\left\{ \varphi \right\} $ are the phase variables defined on the
sites, the sum is over nearest neighbors, $J>0$ is the coupling constant
and $A_{ij}=\frac{2e}{\hslash c}\int_{i}^{j}A{\cdot }dl$ is the line
integral along the bond between adjacent sites $i$ and $j$. We consider the
case where the bond variables $A_{ij}$ are fixed, uniformly quenched, out of
equilibrium with the site variables and satisfy the condition $%
\sum_{p}A_{ij}=2\pi f$; here the sum is over each set of bonds of an
elementary plaquette and $f$ is the strength of frustration. We assume that
the local magnetic field in Eq. (\ref{act1}) is equal to the uniform applied
field; such an approximation is more valid the smaller is the sample size $L$
compared with the transverse penetration depth $\lambda _{\perp }$ \cite
{halperin1}. In the case of full frustration, i.e. $f=\frac{1}{2}$, of interest to
us here, such a model has a continuous $U(1)$ symmetry associated with the
rotation of spins and an extra discrete $Z_{2}$ symmetry, as it has been
shown analysing the degeneracy of the ground state \cite{halsey2,villain}.
Choosing the Landau gauge, such that the vector potential vanishes on all
horizontal bonds and on alternating vertical bonds, we get a lattice where
each plaquette displays one antiferromagnetic and three ferromagnetic bonds.
Such a choice corresponds to switching the sign of the interaction term in
Eq. (\ref{act1}) and is closely related to the presence of two ground states
with opposite chiralities, the first one invariant under shifts by two
lattice spacings and the second one invariant under shifts by one
lattice spacing.

In the presence of such a degeneracy we can use the $m$-reduction technique
\cite{cgm4} which has been successfully applied to a quantum Hall fluid in
\cite{cgm2,cgm4,cgm1,noi1,noi2} and to a fully frustrated Josephson junction
ladder with Mobius boundary conditions in \cite{FFXY-ladders}. In Section 4
we will see how the discrete version of the $m$-reduction procedure,
successfully used to build a CFT with a $Z_{m}$-twist, well describes such a
situation.

The $Z_{2}$ symmetry of the FFXY model is broken at low temperature and will be
restored beyond a certain temperature after the formation of domain walls
separating islands of opposite chirality. The Ising transition overlaps to a vortex-unbinding transition \cite{ktou}, which is associated with the $U(1)$ symmetry \cite{villain,teitel2}.

The action (\ref{act1}) can be cast into a form where both the $U(1)$ and $%
Z_{2}$ symmetries are manifest, through the Villain approximation \cite
{villain2}. In this way the spin-wave and the vortex contributions can be
separated. Furthermore, by integrating out the spin waves, the resulting
vortex contribution can be rewritten as a fractionally charged Coulomb gas
(CG) defined on the dual lattice \cite{villain,fradkin}:
\begin{equation}
H=-\frac{J}{kT}\sum_{r,r^{\prime }}\left( m\left( r\right) +f\right) G\left(
r,r^{\prime }\right) \left( m\left( r^{\prime }\right) +f\right) .
\label{act2}
\end{equation}
Here we have $\lim_{\left| r-r^{\prime }\right| \rightarrow \infty }G\left(
r,r^{\prime }\right) =\log \left| r-r^{\prime }\right| +\frac{1}{2}\pi $ and
the neutrality condition $\sum_{r}\left( m\left( r\right) +f\right)
=\sum_{r}n\left( r\right) =0$ must be satisfied. It is now evident that the
ground state for $f=\frac{1}{2}$ consists of an alternating lattice of
logarithmically interacting $\pm \frac{1}{2}$ charges and is doubly
degenerate. Such a model exhibits two possible phase transitions, an Ising
and a vortex-unbinding one, and their relative order has been deeply studied
in the literature \cite{halsey2}.

It is just the doubly degenerate checkerboard pattern of vortices
in the ground state that gives rise to the Ising-like $Z_{2}$
discrete symmetry in addition to the Kosterlitz-Thouless (KT)
continuous $U(1)$ one, associated with the uniform rotation of all
phase angles \cite{ktou}. The issue whether
there are two distinct phase transitions, $T_{V}>T_{ISING}$ or $T_{V}<T_{ISING}$ with $T_{V}$ and $%
T_{ISING}$ marking respectively the breaking of $U(1)$ and of
$Z_{2}$ symmetry \cite{double}, or a single transition with the
simultaneous breaking of both symmetries \cite{single} has been
widely investigated. Indeed an exact mapping of the classical 2D
FFXY model onto an alternating $19$-vertex model has been
established\cite{quantum1}, whose equivalence with the restricted
solid-on-solid (RSOS) model is well known\cite{quantum2}. In turn
the RSOS model can be viewed as a discretized time generalization
of the spin-$1$ one dimensional quantum chain, as shown in detail
in Ref.\cite{quantum3} by means of the transfer matrix formalism.
The phase diagram of such a system has been widely investigated in
a simplified case, i.e. the XXZ quantum chain with single-ion-type
anisotropy, by means of analytical and numerical techniques
\cite{quantum3,quantum4,quantum5}. It shows both $U(1)$ and
$Z_{2}$ symmetry breaking in the space $D-J_{z}$ (where the
parameter $D$ is the uniaxial single-ion anisotropy)
\cite{quantum3,quantum6,quantum7}. Depending on the values of the
parameters different phases of the system emerge: the Haldane
phase, the large-$D$ phase, two $XY$ phases ($XY1$ and $XY2$) with
different $Z_{2}$ symmetry, the ferromagnetic phase and the Neel
phase. Then various types of phase transitions take place. In
particular all the three possibilities above quoted can be
recognized: a gapful-gapless KT transition between the Haldane
phase and the $XY1$ one followed by a transition between the two
different $XY$ phases, $XY1$ and $XY2$; an Ising transition
between the Haldane phase and Neel phase and after that a
transition Neel-$XY2$ at $J_{z}=0$ for large negative $D$; an $XY$
and Ising transition lines which merge at a tricritical point
within $c=\frac{3}{2}$ CFT universality class. All that reveals
the richness of the phase structure of a class of frustrated
systems with the same ground state but the controversy on the
phase diagram structure of the $FFXY$ model remains still
unsolved.

Recently Korshunov \cite{korsh} argued that a new transition could
take place in the 2D FFXY model, well below the bulk transition,
due to the unbinding of kink-antikink pairs on the domain walls
associated with $Z_{2}$ symmetry. Such a transition could lead to
a decoupling of phase coherence across
domain boundaries, so producing two distinct bulk transitions with $%
T_{V}<T_{ISING}$ and then a coupled XY-Ising model
\cite{granato,blote}. Direct numerical evidence supporting the
existence of such a transition at a temperature $T_{W}$ below the
bulk ones has by now been provided \cite {olsson}. Also Teitel and
Jayaprakash \cite{teitel2} have considered the case
$T_{V}<T_{ISING}$ in terms of a dual Coulomb gas model. They
argued that the helicity modulus $Y$ would decrease from its $T=0$
value by means of fluctuations producing dipole moments. So, if
Ising domains of size $\xi _{ISING}$ carried a total dipole moment
proportional to their size, they would make $Y=0$ in a continuous
way, due to the divergence of $\xi _{ISING}$ as $T\rightarrow
T_{ISING}$. At $T=0$ it has been found though that the only
Ising-like domains are those which carry no dipole moment: they
cannot produce a reduction in $Y$. But at $T>T_{W}$, according to
Ref. \cite{korsh}, kink-antikink excitations on the boundaries of
the domain walls unbind, so the Ising domains would acquire large
dipole moments without cost in free energy. Such a conclusion
\cite {olsson} supports strongly the existence of two separate
transitions with $T_{V}<T_{ISING}$. This scenario has been
recently confirmed also by numerical simulations on very large
lattices and by means of a careful finite-size scaling analysis in
Refs. \cite{vicari}. The critical behaviour at the overlap
temperature is well described by a conformal field theory with
central charge $c=3/2$ \cite{foda}. The interesting scenario just
outlined is the result of an interplay between the Ising and
gaussian sectors of the model: they are linked through the
fractional vortices which reside at the corners of the domain
walls in the off-critical theory at the microscopic level
\cite{halsey2}. More recent Monte Carlo simulations of $2D$ $XY$-type models
\cite{petter1} appear to confirm the above scenario
\cite{petter2}. Indeed the sequence of phase transitions on going
from higher to lower temperatures is: first Ising, then KT and
next kink-antikink unbinding. This picture will then remain (with
Ising and KT transition very close to each other) until a
particular value of the parameters where all the three lines merge
at the same time. After that the transition is first order and the
critical point is characterized by a central charge $c=2$
\cite{petter2}.

The FFXY model has an interesting experimental realization as a
Josephson junctions array (JJA) in a transverse magnetic field
with half a magnetic flux quantum per cell \cite{mooij,lobb}.
Measurements have been performed on the electrical properties of
such a system, in particular on the resistance as a function of
the temperature and the external magnetic field which is closely
related to the helicity modulus and then to the vortex-unbinding
transition \cite{mooij}. The fact that the vortex transition is
shifted downwards in temperature \cite{mooij} is in agreement with
the results found in Ref. \cite{sokalski}, which refer to an XY
model with a modified potential of the form:
\begin{equation}
V\left( \varphi \right) =\cos \varphi +\alpha \cos 2\varphi \text{.}
\end{equation}
That tentatively leads to the conclusion that the arrays of Ref. \cite{mooij}
display all the properties of the FFXY model with higher harmonics
contributions to the potential \cite{foda}.

In this paper we construct a CFT for such systems which extends the results
of Ref. \cite{foda}, so recovering the whole phase diagram quoted in Refs.
\cite{granato,blote,granato0,baseilhac}, then we close with a brief account
on the relation between the FFXY model and the physics of a Josephson
junction array in the classical approximation \cite{fazio}.

The paper is organized as follows.

In Section 2 we discuss in a qualitative way the phase diagram of the FFXY
model describing some models which have the same $U(1)\otimes Z_{2}$
degenerate ground state and are believed to be in the same universality
class.

In Section 3 we recall some aspects of the $m$-reduction procedure, in
particular we show how the $m=2$, $p=0$ case well accounts for the $%
U(1)\otimes Z_{2}$ symmetry of the FFXY model. In such a framework we give
the whole primary fields content of the theory on the plane.

In Section 4 we give a discrete version of the $m$-reduction procedure and
then, starting from our CFT results, we show how to recover the whole FFXY
phase diagram introduced in Section 2.

In Section 5 some comments and outlooks are given.

\section{Phase diagram of the FFXY model}

In the literature the phase transitions of the FFXY model have been studied
analyzing other models which have the same $U(1)\otimes Z_{2}$ degenerate
ground state and are believed to be in the same universality class. In
particular, by using symmetry arguments\cite{halsey2,single} or a
Hubbard-Stratonovich transformation\cite{granato}, the FFXY model can be
reformulated in terms of a system of two coupled XY models with a symmetry
breaking term:
\begin{equation}
H=A\left[ \sum_{i=1,2}\sum_{\left\langle r,r^{^{\prime }}\right\rangle }\cos
\left( \varphi ^{(i)}(r)-\varphi ^{(i)}(r^{^{\prime }})\right) \right]
+h\sum_{r}\cos 2\left( \varphi ^{(1)}(r)-\varphi ^{(2)}(r)\right) .
\label{H-GK}
\end{equation}
Such a system has been studied in detail by Granato and Kosterlitz \cite
{granato1}. The limit $h\rightarrow 0$ corresponds to a full decoupling of
the fields $\varphi ^{(i)}$, $i=1,2$, so giving rise to a CFT with central
charge $c=2$ which describes two independent classical XY models (in the
continuum limit). In the $h\rightarrow \infty $ limit, the two phases $%
\varphi ^{(i)}$, $i=1,2$ are locked \cite{granato1}:
\begin{equation}
\varphi ^{(1)}(r)-\varphi ^{(2)}(r)=\pi j \text{, \ \ \ \ }j=1,2;
\end{equation}
as a consequence the model gains a symmetry $U(1)\otimes Z_{2}$ and its
Hamiltonian renormalizes towards a model described by:
\begin{equation}
H=H\left( h\rightarrow \infty \right) =A\sum_{\left\langle r,r^{^{\prime
}}\right\rangle }\left( 1+s_{r}s_{r^{^{\prime }}}\right) \cos \left( \varphi
^{(1)}(r)-\varphi ^{(1)}(r^{^{\prime }})\right)
\end{equation}
where $\varphi ^{(1)}(r)$ and $s_{r}=\cos \pi j=\pm 1$ are planar and
Ising spins respectively. In this way a model is obtained which is
consistent with the required symmetry, the XY-Ising one and whose
Hamiltonian has the general form \cite{granato,granato0}:
\begin{equation}
H_{XY-I}=\sum_{\left\langle r,r^{^{\prime }}\right\rangle }\left[ A\left(
1+s_{r}s_{r^{^{\prime }}}\right) \cos \left( \varphi ^{(1)}(r)-\varphi
^{(1)}(r^{^{\prime }})\right) +Cs_{r}s_{r^{^{\prime }}}\right] ,
\label{hamil4}
\end{equation}
which is believed to be in the universality class of the classical FFXY
model. However such a model does not contain fractional vortices at the
corners of the domain walls \cite{halsey2}, which are believed to give rise
to an interesting interplay between the continuous and discrete symmetries
\cite{foda}. It is just this complex interplay which accounts for the
unusual and not yet fully understood critical behaviour of the FFXY model on
the square lattice \cite{foda,granato,blote,granato0,baseilhac}, that will
be discussed in the following.

The phase diagram of the model \cite{granato,blote,granato0,baseilhac} with
Hamiltonian (\ref{hamil4}), as depicted in Fig. 1, is built up with three
branches which meet at a multi-critical point $P$. The branch $PT$
corresponds to single transitions with simultaneous loss of XY and Ising
order while the other two describe separate Kosterlitz-Thouless and Ising
transitions. The critical line $PT$ becomes a first order one at the
tricritical point $P$; it seems to be non-universal \cite{granato,minn} as
the values of the critical exponents ($\nu =0.79$, $\eta =0.40$)
estimated by finite-size scaling of large systems are inconsistent with pure
Ising critical behavior ($\nu =1$, $\eta =\frac{1}{4}$). Also the numerical
estimate for the central charge $c$ ($c\sim 1.60$) \cite{blote} is higher
than the expected value ($c=3/2$) for a critical behavior given by the
superposition of critical Ising ($c=1/2$) and gaussian ($c=1$) models \cite
{foda}. Indeed the central charge seems to vary continuously from $c\approx
1.5$ near $P$ to $c\approx 2$ at $T$ \cite{blote}. All such results would
suggest the existence of a parameter changing along the critical $PT$ line
that does not affect the symmetry \cite{baseilhac}. The system seems to be
not conformally invariant \cite{granato}, so it would be natural to consider
all possible integrable perturbations of a superposition of critical Ising
and XY models as a starting point to study the vicinity of the point $P$ \cite
{baseilhac}. Furthermore different numerical techniques have led to
different results suggesting either single \cite{single} or double
transitions \cite{double} along the critical $PT$ line. How to reconcile
such conflicting results? A possible solution is suggested in Ref. \cite
{knops1} where the Coulomb gas Hamiltonian of Eq. (\ref{act2}) gets modified
through an additional antiferromagnetic coupling between nearest-neighbour
vortices:
\begin{equation}
H=-\frac{J}{kT}\sum_{r,r^{^{\prime }}}n_{r}G\left( r,r^{^{\prime }}\right)
n_{r^{^{\prime }}}+\overline{J}\sum_{\left\langle r,r^{^{\prime
}}\right\rangle }n_{r}n_{r^{^{\prime }}}.  \label{ham5}
\end{equation}

\begin{figure}[tbp]
\centering \includegraphics*[width=0.9\linewidth]{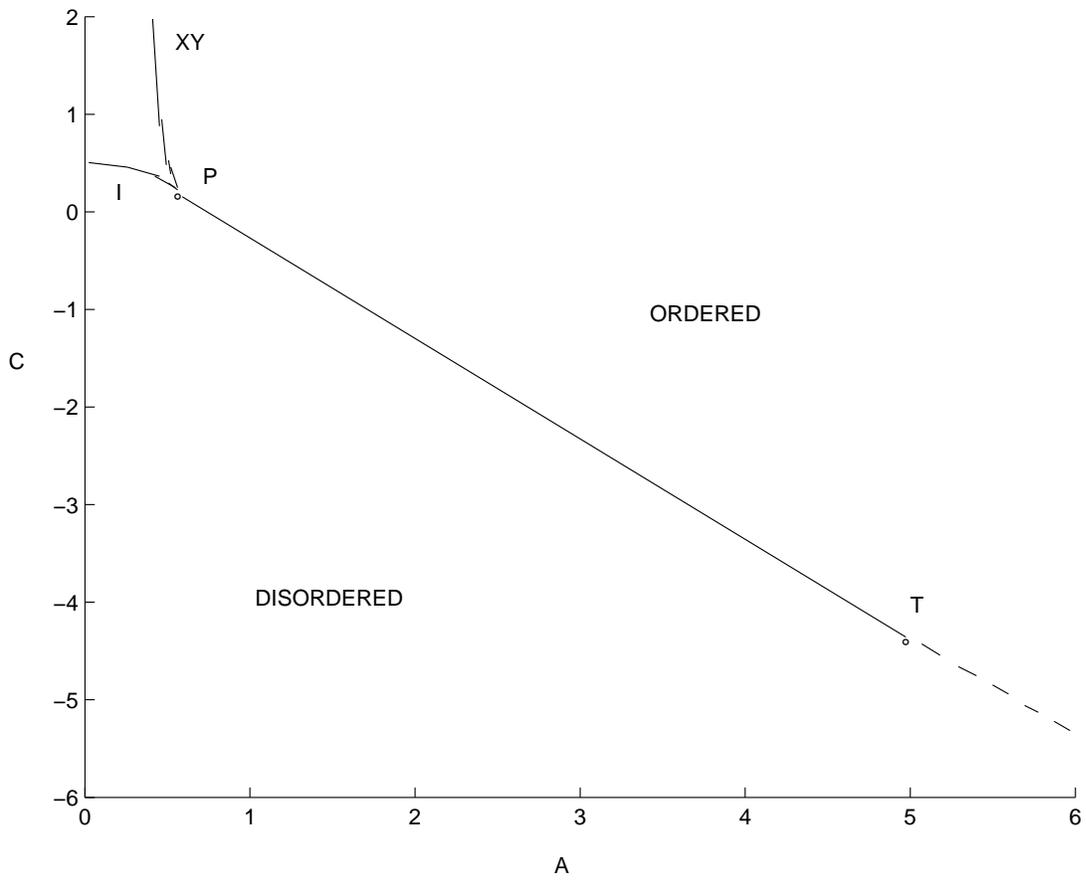}
\caption{The phase diagram for the XY-Ising model \protect\cite
{granato,granato0}. }
\label{figura1}
\end{figure}

Such a model has a phase diagram with a structure close to the XY-Ising
model one of Fig. 1 with the occurrence of double or single transitions on
the $PT$ line depending on this additional coupling; for $\overline{J}\neq 0$
the two transitions separate with the vortex-unbinding one occurring at a
lower temperature \cite{knops1}. A similar effect has been observed
theoretically in Ref. \cite{sokalski}, where a XY model with various
modifications of the cosine-type potential is studied, and experimentally in
Ref. \cite{mooij}: it can be ascribed to the presence of higher harmonics
contributions to the potential. Such a finding is a crucial one, as we will
show in Section 4.

Another route to the solution of all these issues is the $m$-reduction
technique \cite{cgm2,cgm4} which describes well the two fixed points with $%
c=3/2$ and $c=2$ and accounts for non trivial boundary conditions for the
square lattice. Such a realization of the FFXY model on closed geometries
could be relevant for the description of JJAs with non trivial topologies,
which are believed to provide a physical implementation of an ideal quantum
computer \cite{ioffe} because of the topological ground state degeneracy
``protected''\ from external perturbations \cite{wen}. We introduce the $m$%
-reduction in the next Section.

\section{A review of the $m$-reduction procedure}

In this Section we recall those aspects of the twisted model (TM) which are
relevant for the FFXY model. We focus in particular on the $m$-reduction
procedure for the special $m=2$ case (see Ref. \cite{cgm2} for the general
case), since we are interested in a system with $U(1)\otimes Z_{2}$
symmetry. We showed in Refs. \cite{cgm2,cgm4} that such a theory describes
well a system consisting of two parallel layers of 2D electron gas in a
strong perpendicular magnetic field, with filling factor $\nu ^{(a)}=\frac{1%
}{2p+2}$ for each of the two $a=1$, $2$ layers and total filling $\nu =\nu
^{(1)}+\nu ^{(2)}=\frac{1}{p+1}$. Regarding the integer $p$, characterizing
the flux attached to the particles, we choose the ``bosonic''\ value $p=0$,
since it enables us to describe the highly correlated system of vortices
with flux quanta $\frac{hc}{2e}$.

Let us start from the bosonic ``filling'' $\nu =\frac{1}{2}$, described by a
CFT with $c=1$ in terms of a scalar chiral field $Q$ compactified on a
circle with radius $R^{2}=1/\nu =2$. It is explicitly given by:
\begin{equation}
Q(z)=q-i\,p\,lnz+i\sum_{n\neq 0}\frac{a_{n}}{n}z^{-n}\text{,}  \label{modes}
\end{equation}
with $a_{n}$, $q$ and $p$ satisfying the commutation relations $\left[
a_{n},a_{n^{\prime }}\right] =n\delta _{n,n^{\prime }}$ and $\left[ q,p%
\right] =i$.

From such a CFT (mother theory), using the $m$-reduction procedure, which
consists in considering the subalgebra generated only by the modes in Eq. (%
\ref{modes}) which are a multiple of an integer $m$, we get a $c=m$ orbifold
CFT (daughter theory, i.e. the TM) which, for $m=2$, in the next Section
will be shown to describe the whole phase diagram of the FFXY model. Then
the fields in the mother CFT can be organized into components which have
well defined transformation properties under the discrete $Z_{m}$ (twist)
group, which is a symmetry of the TM. By using the mapping $z\rightarrow
z^{1/m}$ and by making the identifications $a_{nm+l}\longrightarrow \sqrt{m}%
a_{n+l/m}$, $q\longrightarrow \frac{1}{\sqrt{m}}q$, the $c=m$ CFT (daughter
theory) is obtained.

Its primary fields content, for the special $m=2$ case, can be expressed in
terms of a $Z_{2}$-invariant scalar field $X(z)$, given by
\begin{equation}
X(z)=\frac{1}{2}\left( Q(z)+Q(-z)\right) ,  \label{X}
\end{equation}
describing the electrically ``charged'' sector of the new theory, and a
twisted field
\begin{equation}
\phi (z)=\frac{1}{2}\left( Q(z)-Q(-z)\right) ,  \label{phi}
\end{equation}
which satisfies the twisted boundary conditions $\phi (e^{i\pi }z)=-\phi (z)$
and describes the ``neutral'' sector \cite{cgm2}.

The TM primary fields are composite vertex operators $V\left( z\right)
=U_{X}\left( z\right) \psi \left( z\right) $ where
\begin{equation}
U_{X}\left( z\right) =\frac{1}{\sqrt{z}}:e^{i\alpha X(z)}:\text{, \ \ \ \ }%
\alpha ^{2}=2  \label{char}
\end{equation}
is the vertex of the ``charged''\ sector and
\begin{equation}
\psi \left( z\right) =\frac{1}{\sqrt{z}}:e^{i\alpha \phi \left( z\right) }:
\label{neu}
\end{equation}
is the ``neutral''\ one. We point out that in the neutral sector it is useful to introduce the two chiral operators:
\begin{equation}
\psi \left( z\right) =\frac{1}{2\sqrt{z}}\left( :e^{i\alpha \phi \left(
z\right) }:+:e^{i\alpha \phi \left( -z\right) }:\right) ,\text{ \ \ }%
\overline{\psi }\left( z\right) =\frac{1}{2\sqrt{z}}\left( :e^{i\alpha \phi
\left( z\right) }:-:e^{i\alpha \phi \left( -z\right) }:\right) ;
\label{neu1}
\end{equation}
the first one does not change the boundary conditions while the second one
does. In a fermionized version of the theory they correspond to two $c=1/2$
Majorana fermions with Ramond and Neveu-Schwartz boundary conditions \cite
{cgm2,cgm4}. Furthermore in the TM they appear to be not completely
equivalent.

In fact the whole TM theory decomposes into a tensor product of two CFTs, a
twisted invariant one with $c=3/2$ (the Moore-Read (MR) theory with symmetry
$U(1)\otimes Z_{2}$) and the remaining $c=1/2$ one realized by a Majorana
fermion in the twisted sector. Such a factorization can be unambiguously
pointed out on the torus topology where the partition function of the $c=2$
theory, i.e. the TM, can be written as\footnote{%
See Appendix for the explicit expression and Ref. \cite{cgm4} for its
derivation.}:
\begin{equation}
Z\left( w_{c}|\tau \right) =Z^{MR}\left( w_{c}|\tau \right) Z_{\overline{I}%
}\left( \tau \right) .  \label{fullpf}
\end{equation}
Here $Z^{MR}\left( w_{c}|\tau \right) $ stands for the MR partition function
and $Z_{\overline{I}}\left( \tau \right) $ is the partition function
describing the Ising degrees of freedom which are $Z_{2}$ antisymmetric (i.e. $\overline{\psi }\left( z\right) $ of Eq. (\ref{neu1}) in the plane
geometry).

Let us point out that the energy-momentum tensor of the Ramond part of the
neutral sector develops a cosine term:
\begin{equation}
T_{\psi }\left( z\right) =-\frac{1}{4}\left( \partial \phi \right) ^{2}-%
\frac{1}{16z^{2}}\cos \left( 2\sqrt{2}\phi \right) ,
\end{equation}
a clear signature of a tunneling phenomenon which selects out a new stable
vacuum, the $c=3/2$ one. We identify such a $Z_{2}$ invariant theory with
the one describing the FFXY model conjectured in Ref. \cite{foda}. Here we
only point out that the critical behavior of the $c=3/2$ theory will be a
superposition of the critical behavior of its components with $c=1$ and $%
c=1/2$ respectively. The gaussian component will lead to a gaussian critical
line that ends at a vortex-unbinding transition point, while the Ising component
will lead to an Ising critical point. The conformal field theory, considered
in this paper, is not meant to predict the position of the Ising point with
respect to the gaussian critical line, it only tells us that there exists an
Ising transition overlapping somewhere with a gaussian critical line. In the
following we exhibit all the relevant physical consequences of our theory.

\section{TM description of the FFXY phase diagram}

In this Section we will derive the phase diagram of the FFXY model\footnote{%
As it has been described in Section 2 and inferred by Migdal-Kadanoff
renormalization \cite{granato0}, Monte Carlo simulations \cite{granato} and
Monte Carlo transfer matrix calculations \cite{blote}.} in terms of the RG
flow which originates from perturbing our TM model.

As we pointed out in Section 2, the limit $h\rightarrow 0$ in the
Hamiltonian of Eq. $\left( \ref{H-GK}\right) $ corresponds to a full
decoupling of the fields $\varphi ^{(i)}$, $i=1,2$. In the continuum limit
that gives rise to a CFT with two scalar Fubini boson fields $\varphi ^{(i)}$
and with central charge $c=2$. Let us stress that a good candidate to
describe the FFXY model at criticality around the point T of the phase
diagram is a CFT, with $c=2$, which accounts for the full spectrum of
excitations of the model. So it is worth recalling briefly the structure of
such excitations for the FFXY model. A domain wall is a topological
excitation of the double degenerate ground state and it can be defined as a
line of links, each one separating two plaquettes with the same chirality.
So, through a domain wall the alternating structure (the checkerboard
pattern) of the ground state is lost. Kinks and antikinks are excitations
which live on the domain walls and are described by fractional vortices with
+1/2 and $-1/2$ topological charge. Such particle-like excitations belong
also to the spectrum of the fully frustrated ladders of Josephson junctions
and in \cite{FFXY-ladders} we have shown how the TM model exactly describes
them. In particular these excitations can be generated from the ground state
by closing the ladder, by imposing the coincidence of opposite sides. In
fact, in this closed geometry, for an even number of plaquettes we obtain
the ground state while for the odd case this point-like excitation is
generated on the boundary. In the TM model they are described by different
boundary conditions, untwisted and twisted ones, respectively.

Similar considerations apply to the FFXY array; in particular the central
role played by the closed geometry and by the boundary conditions in the
description of the excitation spectrum\ is clear already at the level of the
lattice model. By imposing the coincidence between opposite sides of the
square lattice, we obtain a closed geometry, which is the discretized
analogue of a torus. Now for the ground state two topologically inequivalent
circumstances arise, one for even and the other one for odd number of
plaquettes. In the even case the end plaquettes on the opposite sides of the
lattice have opposite chirality, while in the odd case they have the same
chirality. So the ground state on the square lattice maps into the ground
state for the even case while it generates two straight domain walls along
the two cycles of the torus for the odd case. Such a behaviour has to be
taken into account by opportune boundary conditions on the field $\varphi
^{(i)}$ at the borders of the finite lattice. These non trivial boundary
conditions naturally arise when we implement the $m$-reduction procedure in the discrete case. To this aim, let $(-L/2,0),$ $(L/2,0)$, $(L/2,L)$, $%
(-L/2,L)$ be the corners of the square lattice ${\cal L}$ and assume that
the fields $\varphi ^{(i)}$ satisfy the following boundary conditions:
\begin{equation}
\varphi ^{(1)}(r)=\varphi ^{(2)}(r)\text{\ \ \ for }r\in {\cal L}\cap {\bf x}%
,
\end{equation}
where ${\bf x}$ is the $x$ axis. The above boundary conditions allow us to
consider the two fields $\varphi ^{(1)}$ and $\varphi ^{(2)}$ on the square
lattice ${\cal L}$ as the folding of a single field ${\cal Q}$, defined on
the lattice ${\cal L}_{0}$ with corners $(-L/2,-L),$ $(L/2,-L)$, $(L/2,L)$, $%
(-L/2,L)$. More precisely we define the field ${\cal Q}$ as:
\begin{equation}
{\cal Q}(r)=\left\{
\begin{array}{c}
\varphi ^{(1)}(r)\text{ \ \ \ \ \ \ \ \ \ \ for }r\in {\cal L}\cap {\cal L}%
_{0}, \\
\varphi ^{(2)}(-r)\text{ \ \ \ for }r\in (-{\cal L)}\cap {\cal L}_{0}.
\end{array}
\right.
\end{equation}
We can implement now a discrete version of the $m$-reduction procedure ($m=2$%
) by defining the fields:
\begin{equation}
{\cal X}(r)=\frac{1}{2}\left( {\cal Q}(r)+{\cal Q}(-r)\right) ,
\label{X-scalar}
\end{equation}
\begin{equation}
\Phi (r)=\frac{1}{2}\left( {\cal Q}(r)-{\cal Q}(-r)\right) ,
\label{ph-scalar}
\end{equation}
where $r\in {\cal L}_{0}$. The resemblance with the continuum version of the
$m$-reduction procedure is evident and the fields ${\cal X}$ and $\Phi $ are
symmetric and antisymmetric with respect to the action of the generator $%
g:r\rightarrow -r$ of the discrete group $Z_{2}$. The Hamiltonian in Eq. $%
\left( \ref{H-GK}\right) $ can be rewritten in terms of these fields and,
for $h=0$, it becomes:
\begin{equation}
H=2A\sum_{\left\langle r,r^{^{\prime }}\right\rangle \in {\cal L}}\cos
\left( {\cal X}(r)-{\cal X}(r^{^{\prime }})\right) \cos \left( \Phi (r)-\Phi
(r^{^{\prime }})\right) ,  \label{dham2}
\end{equation}
which in the continuum limit corresponds to the action of our TM model:
\begin{equation}
{\cal A}=\int \left[ \frac{1}{2}\left( \partial {\cal X}\right) ^{2}+\frac{1%
}{2}\left( \partial \Phi \right) ^{2}\right] d^{2}x\text{.}
\end{equation}
It is worth pointing out that the fields ${\cal X}$ and $\Phi $ are scalar
fields and so the chiral fields defined by Eqs. (\ref{X}), (\ref{phi}) can
be seen as their chiral components. Moreover the group $Z_{2}$ is a discrete
symmetry group, indeed both $H$ and ${\cal A}$ are invariant under its
action.

In the following we will show how the continuum version of the FFXY model
and its phase diagram can be described by the action:
\begin{equation}
{\cal A}=\int \left[ \frac{1}{2}\left( \partial {\cal X}\right) ^{2}+\frac{1%
}{2}\left( \partial \Phi \right) ^{2}+\mu \cos \left( \beta \Phi \right)
+\lambda \cos \left( \frac{\beta }{2}\Phi +\delta \right) \right] d^{2}x%
\text{,}  \label{X-two-SG}
\end{equation}
defined in terms of $m=2$ reduced fields ${\cal X}$ and $\Phi $, which
embodies in\ its ``neutral'' sector the higher harmonic potential term
conjectured in Refs. \cite{foda,sokalski,knops1}. We assume the constraints $%
\beta ^{2}<8\pi $, which characterizes both the cosine terms as relevant
scalar fields perturbing the gaussian fixed point, and $\left| \delta
\right| \leq \pi /2$ (see \cite{mussardo}). Thus the ``neutral''\ sector is
a two-frequency sine-Gordon theory that can be viewed as a deformation of a
pure sine-Gordon one with the perturbing term $\lambda \cos \left( \beta
\Phi /2+\delta \right) $. The ultraviolet (UV) fixed point $\mu =0,$ $%
\lambda =0$ of the action (\ref{X-two-SG}) corresponds to the TM model with
central charge $c=2$, describing the fixed point T in the phase diagram of
the FFXY model. Here the argument described in \cite{mussardo} can be
adapted to our particular case to study the RG flow in the ``neutral''\
sector of our theory. Let us define the dimensionless variable $\eta \equiv
\lambda \mu ^{-\left( 8\pi -(\beta /2)^{2}\right) /\left( 8\pi -\beta
^{2}\right) }$; when $\eta =0$ ($\lambda =0$) the ``neutral''\ sector
reduces to a sine-Gordon model with a particle spectrum which consists of
solitons and antisolitons and, for $\beta ^{2}<4\pi $, some breathers.
Switching on the perturbation (i.e. for $\lambda \neq 0$) a confinement of
solitons into states with zero topological charge takes place and packets
formed by $2$ of the original solitons or antisolitons survive as stable
excitations for generic values of $\left| \delta \right| <\pi /2$. In the
limit $\eta \rightarrow \infty $ the 2-soliton evolves into the $1$-soliton
of the pure sine-Gordon model with $\mu =0$. An unbinding phenomenon takes
place in the particular $\delta =\pm \pi /2$ case for finite $\eta $ and the
$2$-soliton decomposes into a sequence of two kinks $K_{1}$. In such a case
the limit $\eta \rightarrow \infty $ implies a transmutation of a composite
topological excitation (the two kinks $K_{1}$) into an elementary one (the
1-soliton). So the existence of an intermediate critical value $\eta =\eta
_{c}$ is required at which a phase transition takes place and the RG flow
ends into the infrared (IR) fixed point described by a CFT with central
charge $c=1/2$, the Ising model. The central charge of the full model (\ref
{X-two-SG}) so changes from $c=2$ of the UV fixed point to $c=3/2$ of the IR
fixed point, i.e. we recover early known results of Monte Carlo simulations
\cite{blote}. Indeed in the following we clarify how such an IR fixed point
coincides with the $U(1)\otimes Z_{2}$ symmetric component ($c=3/2$) of the
whole TM model, which results then to properly describe the FFXY model
conjectured in Ref. \cite{foda}, i.e. the fixed point $P$ in the phase
diagram of the FFXY model. Then the full phase diagram of the FFXY model in
Fig. 1 would be recovered.

To such an extent we define fermion fields on two parallel
layers in terms of the ``neutral'' boson fields. In particular let $\psi
_{i}\left( x\right) $ be the fermion on layer $i$, whose chiral
components at the UV fixed point are defined by:
\begin{equation}
\left( \psi _{i}\right) _{L/R}=\frac{1}{\sqrt{z}}:e^{i\alpha \left( \Phi
\right) _{L/R}\left( (-)^{i+1}z\right) }:\text{;}
\end{equation}
where $\left( \Phi \right) _{L/R}$ are the left and the right components of $%
\Phi $. In terms of them the $Z_{2}$ symmetric $\psi $ and $Z_{2}$
antisymmetric $\overline{\psi }$ fermions read as:
\begin{equation}
\psi =\frac{1}{2}(\psi _{1}+\psi _{2})\text{ \ , \ }\overline{\psi }=\frac{1%
}{2}(\psi _{1}-\psi _{2})\text{.}
\end{equation}
The action (\ref{X-two-SG}) at $\left| \delta \right| =\pi /2$ can then be
rephrased in terms of two interacting Ising layers as:
\begin{eqnarray}
{\cal A} &=&\int \frac{1}{2}\left( \partial {\cal X}\right) ^{2}d^{2}x+\,%
{\cal A}_{1}^{Ising}+{\cal A}_{2}^{Ising}+\rho \int d^{2}x(\varepsilon
_{1}\left( x\right) \varepsilon _{2}\left( x\right) )+ \\
&&+\mu \int d^{2}x\left( \varepsilon _{1}\left( x\right) +\varepsilon
_{2}\left( x\right) \right) +\lambda \int d^{2}x(\sigma _{1}(x)\sigma
_{2}(x)),  \label{ham7}
\end{eqnarray}
where ${\cal A}_{i}^{Ising}$ denotes the fixed point action of the Majorana
fermion $\psi _{i}$ and $\varepsilon _{i}\left( x\right) ,\sigma _{i}\left(
x\right) $ are the energy and spin of the $i$-th Ising model ($i=1,2$). This
correspondence comes out observing that the two critical Ising models plus
the interaction term in the coupling $\rho $ define a line of $c=1$ CFTs
parameterized by $\rho $ since the operator $\varepsilon _{1}\varepsilon _{2}
$ is marginal ($\Delta _{\varepsilon _{1}\varepsilon _{2}}=1$). In this
one-parameter family of $c=1$ CFTs the thermal $\varepsilon _{1}+\varepsilon
_{2}$ and the magnetic $\sigma _{1}\sigma _{2}$ operators are known \cite
{teller} to have $\rho $-dependent conformal dimensions which however
satisfy the relation:
\begin{equation}
\frac{\Delta _{\sigma _{1}\sigma _{2}}(\rho )}{\Delta _{\varepsilon
_{1}+\varepsilon _{2}}(\rho )}=\frac{\Delta _{\sigma _{1}\sigma _{2}}(0)}{%
\Delta _{\varepsilon _{1}+\varepsilon _{2}}(0)}=\frac{1}{4}\text{.}
\end{equation}
It allows us to say that the thermal and magnetic terms in (\ref{ham7}) play
the role of the two cosine terms in (\ref{X-two-SG}), in particular the
equivalence works under the following identifications:
\begin{equation}
\varepsilon _{1}+\varepsilon _{2}\sim \cos \left( \beta \Phi \right) ,\text{
}\sigma _{1}\sigma _{2}\sim \sin \left( \frac{\beta }{2}\Phi \right) ,
\end{equation}
where the parameters $\rho $ and $\beta $ are linked by the relation $\Delta
_{\varepsilon _{1}+\varepsilon _{2}}(\rho )=\beta ^{2}/8\pi $. Finally we
are in the position to identify unambiguously the IR fixed point $c=3/2$ of
our theory. Indeed for each value of $\rho (\beta )$\ there is a critical
line in the $\mu \lambda $-plane corresponding to the RG flow of our theory
from the UV fixed point $c=2$ to the IR fixed point $c=3/2$; in particular
we can consider the RG flow selected by the limit $\lambda \rightarrow
\infty $. In such a case the magnetic term in (\ref{ham7}) forces the spins $%
\sigma _{1}$ and $\sigma _{2}$ to align everywhere reducing the two Ising
system to a single Ising one. Thus at the IR fixed point the degrees of
freedom of the fermion $\overline{\psi }$ decouple and we are left with the $%
U(1)\otimes Z_{2}$ symmetric component ($c=3/2$) of the whole TM model, i.e.
the MR theory with partition function $Z^{MR}\left( w_{c}|\tau \right) $, as
it can be immediately seen from Eq. (\ref{fullpf}).

\section{Conclusions and outlooks}

In this paper we proposed a $c=2$ CFT description of the FFXY model, which
extends an early proposal by Foda \cite{foda}. A crucial role in such a
construction is played by the $Z_{2}$ ($Z_{m}$ in general) discrete symmetry
built-in in the $2$ ($m$ in general)-reduction technique, which allows for
non trivial topological properties of the vacuum.

By perturbing the $c=2$ CFT with relevant operators, which provide higher
harmonics contribution to the potential, it was shown how the unbinding of
kink-antikink states can give rise to a massless line flow, which fully
reproduces the phase diagram for the FFXY model. As a result of the flow the
non invariant $Z_{2}$ degrees of freedom decouple, the partition function
gets reduced to the Moore-Read one and the degeneracy of the ground state
lost. The attractive infrared fixed point is identified with the $c=\frac{3}{%
2}$ superconformal field theory, that is a free massless field theory of one
boson and one Majorana fermion, which well describes the critical behavior
of the FFXY at the overlap temperature of the Ising and vortex unbinding
transitions.

It is worth noticing that an interesting realization of the FFXY model
physics just outlined can be given in terms of a two dimensional JJA in a
transverse magnetic field \cite{teitel1,teitel2,granato,fazio}. Such a
system is a periodic array of superconducting islands connected by Josephson
links; each island is characterized through the modulus and the phase of the
order parameter. The lowering of the temperature to its critical value
forces each island to become superconductive but the whole array is in a
resistive state as long as the phases have not acquired a long range order.
The array reaches such a global coherence state at a lower temperature and
its behaviour can be well described only in terms of the order parameters
phases $\varphi _{i}$ through the Hamiltonian:
\begin{equation}
H_{JJA}=-E_{J}\sum_{\left\langle ij\right\rangle }\cos \left( \varphi
_{i}-\varphi _{j}\right) ,  \label{ham9}
\end{equation}
where $E_{J}>0$ is the Josephson coupling energy and the sum is over nearest
neighbours. Let us notice the similarity of Eq. (\ref{ham9}) with the
Hamiltonian of a two dimensional XY model. In this picture it is known that
the array becomes superconducting below the temperature $T_{J}=\frac{\pi
E_{J}}{2}$, where the phases are ordered and vortices may appear only in
bounded pairs. Hence the ferromagnetic state for the XY model corresponds to
the superconducting state for the JJA \cite{ktou}. Such a regime is the
classical one for a JJA, indeed the quantum fluctuations of the phases $%
\varphi _{i}$ are weak and the vortices are well defined objects, which form
a Coulomb gas and behave as particles with masses.

The analogy just outlined still works very well in the presence of a
magnetic field perpendicular to the plane of the array, in this case a frustrated XY model can be realized; in particular, the full frustration $f=\frac{1%
}{2}$ corresponds to half a magnetic flux threading each cell. The ground
state can be viewed as a pinned vortex lattice commensurate with the
underlying periodic pinning potential leading to discrete symmetries, which
add to the $U\left( 1\right) $ symmetry of the superconducting order
parameter \cite{teitel1,teitel2}. Furthermore we must remark the striking
role of fully frustrated JJAs as an experimental tool for the study of phase
transitions in FFXY type models \cite{mooij,lobb}.\bigskip

{\bf Acknowledgments.}~~We would like to thank E. Orignac for his careful
reading and comments on a preliminary version of this work. G. N. was
supported by a postdoctoral fellowship of the Minist\`{e}re fran\c{c}ais
d\'{e}l\'{e}gu\'{e} \`{a} l'Enseignement sup\'{e}rieur et \`{a} la Recherche.

\appendix

\section{TM on a torus}

The partition function of the TM model on the torus has the following
factorized form (see \cite{cgm4}):
\begin{equation}
Z(w_{c}|\tau )=Z^{MR}(w_{c}|\tau )Z_{\overline{I}}(\tau )\text{,}
\end{equation}
in terms of the Moore-Read partition function $Z^{MR}$ ($c=3/2$) and of the
Ising partition function $Z_{\overline{I}}$ ($c=1/2$). They have the
following expression:
\begin{equation}
Z^{MR}(w_{c}|\tau )=\left| \chi _{0}^{MC}(0|w_{c}|\tau )\right| ^{2}+\left|
\chi _{1}^{MC}(0|w_{c}|\tau )\right| ^{2}+\left| \chi _{2}^{MC}(0|w_{c}|\tau
)\right| ^{2},
\end{equation}
\begin{equation}
Z_{\overline{I}}(\tau )=\left| \bar{\chi}_{0}(\tau )\right| ^{2}+\left| \bar{%
\chi}_{\frac{1}{2}}(\tau )\right| ^{2}+\left| \bar{\chi}_{\frac{1}{16}}(\tau
)\right| ^{2}.
\end{equation}
The above characters are defined by:
\begin{align}
\bar{\chi}_{0}(\tau )& =\frac{1}{2}\left( \sqrt{\frac{\theta _{3}(0|\tau )}{%
\eta (\tau )}}+\sqrt{\frac{\theta _{4}(0|\tau )}{\eta (\tau )}}\right) , \\
\bar{\chi}_{\frac{1}{2}}(\tau )& =\frac{1}{2}\left( \sqrt{\frac{\theta
_{3}(0|\tau )}{\eta (\tau )}}-\sqrt{\frac{\theta _{4}(0|\tau )}{\eta (\tau )}%
}\right) , \\
\bar{\chi}_{\frac{1}{16}}(\tau )& =\sqrt{\frac{\theta _{2}(0|\tau )}{2\eta
(\tau )}},
\end{align}
which express the primary field content of the Ising model with
Neveu--Schwartz ($Z_{2}$ twisted) boundary conditions, and by:
\begin{eqnarray}
\chi _{0}^{MC}(0|w_{c}|\tau ) &=&\chi _{0}(\tau )K_{0}(w_{c}|\tau )+\chi _{%
\frac{1}{2}}(\tau )K_{2}(w_{c}|\tau )\,,  \label{mr1} \\
\chi _{1}^{MC}(0|w_{c}|\tau ) &=&\chi _{\frac{1}{16}}(\tau )\left(
K_{1}(w_{c}|\tau )+K_{3}(w_{c}|\tau )\right) ,  \label{mr2} \\
\chi _{2}^{MC}(0|w_{c}|\tau ) &=&\chi _{\frac{1}{2}}(\tau )K_{0}(w_{c}|\tau
)+\chi _{0}(0|\tau )K_{2}(w_{c}|\tau )  \label{mr3}
\end{eqnarray}
which express the primary field content of the $Z_{2}$-invariant $c=\frac{3}{%
2}$ \ CFT. They are given in terms of a ``charged''\ contribution ($c=1$):
\begin{equation}
K_{2l+i}(w_{c}|\tau )=\frac{1}{\eta \left( \tau \right) }\;\Theta \left[
\begin{array}{c}
\frac{2l+i}{4} \\[6pt]
0
\end{array}
\right] (2w_{c}|4\tau )\,,\qquad \forall \left( l,i\right) \in \left(
0,1\right) ^{2}\,,  \label{chp}
\end{equation}
and a ``isospin''\ one $\chi _{\beta }(\tau )$ ($c=1/2$), where $w_{c}=%
\dfrac{1}{2\pi i}\,\ln z_{c}$ is the torus variable of the ``charged''\
component.

\end{document}